\documentclass{jfm}
\usepackage{lineno,hyperref}
\usepackage{amsmath}
\usepackage{subfigure}
\usepackage{graphicx}
\usepackage[export]{adjustbox}
\usepackage[table,usenames,dvipsnames]{xcolor}
\usepackage[normalem]{ulem}
\usepackage{graphicx}
\usepackage{epstopdf, epsfig}
\usepackage[ruled,vlined]{algorithm2e}
\usepackage{mathtools}

\def\m1line{\vrule width3pt height2.5pt depth -2pt}

\def\bdot{\raise.2em\hbox to .15em{.}}

\usepackage{todonotes}
\def\outline#1{\textcolor{Gray}{}}
\long\def\comment#1{}

\newcommand{\btau}{{\mbox{\boldmath $\tau$}}}
\newcommand{\bomega}{{\mbox{\boldmath $\omega$}}}
\newcommand{\bx}{{\mbox{\boldmath $x$}}}
\newcommand{\bu}{{\mbox{\boldmath $u$}}}
\newcommand{\ba}{{\mbox{\boldmath $a$}}}
\newcommand{\bn}{{\mbox{\boldmath $n$}}}
\newcommand{\bA}{{\mbox{\boldmath $A$}}}
\newcommand{\bD}{{\mbox{\boldmath $D$}}}
\newcommand{\bX}{{\mbox{\boldmath $X$}}}

\shorttitle{
Origin of Enhanced Skin Friction at the onset of transition
}
\shortauthor{M. Wang, G. L. Eyink \& T. A. Zaki} 

\title{Origin of Enhanced Skin Friction at the Onset of Boundary-Layer Transition}
\author{Mengze Wang\aff{1}, Gregory L. Eyink\aff{1}\aff{,2} \and Tamer A. Zaki\aff{1}\corresp{\email{t.zaki@jhu.edu}}}
\affiliation{\smaller\aff{1}Department of Mechanical Engineering, Johns Hopkins University, Baltimore, MD 21218, USA, 
\aff{2}Department of Applied Math and Statistics, Johns Hopkins University, Baltimore, MD 21218, USA}

\begin{document}
	
\maketitle

\begin{abstract}
Boundary-layer transition is accompanied by a significant increase in skin friction whose origin is rigorously explained using the stochastic Lagrangian formulation of the Navier-Stokes equations. This formulation permits the exact analysis of vorticity dynamics in individual realizations of a viscous incompressible fluid flow.  The Lagrangian reconstruction formula for vorticity is here extended for the first time to Neumann boundary conditions (Lighthill source).  We can thus express the wall vorticity, and therefore the wall stress, as the expectation of a stochastic Cauchy invariant in backward time, with contributions from (a) wall-vorticity flux (Lighthill source) and (b) interior vorticity that has been evolved by nonlinear advection, viscous diffusion, vortex stretching and tilting.
We consider the origin of stress maxima in the transitional region, examining a sufficient number of events to represent the increased skin friction.
The stochastic Cauchy analysis is applied to each event to trace the origin of the wall vorticity.
We find that the Lighthill source, vortex tilting, diffusion and advection of outer vorticity make minor contributions.
They are less important than spanwise stretching of near-wall spanwise vorticity, which is the dominant source of skin-friction increase during laminar-to-turbulent transition.
Our analysis should assist more generally in understanding drag generation and reduction strategies and flow separation in terms of near-wall vorticity dynamics.
\end{abstract}
	
%

\section{Introduction}
\label{sec:intro}

The frictional drag of wall-bounded flows is significantly increased during laminar-to-turbulence transition. 
For example, the friction factor in a smooth pipe increases by more than 50\% as the flow breaks down to turbulence \citep{Moody1944,Mckeon2004}, and the instantaneous wall stresses at the onset of boundary-layer transition can be much higher than in fully turbulent regions 
\citep{Kleiser_Zang_1991,Durbin2007}.
A better understanding of the physical mechanisms driving enhanced skin friction may inform the development of drag-reduction control schemes \citep{Choi1994,Bewley2001} and the interpretation of wall measurements \citep{Wang2021}. 
In this work, we adopt a stochastic Lagrangian approach and provide a precise and quantitative analysis of the origin of the enhanced skin friction in a zero-pressure-gradient boundary layer undergoing bypass transition. 

Since the early stage of fluid mechanics research, it has been speculated that the enhanced frictional drag is related to transitional flow structures.
Based on the analysis of pipe-flow experiments, \cite{Reynolds1883} conjectured that ``...  above this point (critical velocity) the resistance depended upon eddies which might be somewhat uncertain in their action.''
\cite{Lighthill} summarized the theoretical and experimental results for the transitional boundary layer, 
where he noted that turbulence ``concentrates most of the vorticity much closer to the wall than before, although 
at the same time allowing some straggling vorticity to wander away from it farther'',  emphasizing the generality of this process for wall-bounded turbulence and noting further 
that ``during transition..., the mean vorticity at the wall, $\bar{\omega}_w$ (which is 
$\tau_w/\mu,$ where $\tau_w$ is the skin friction), has risen to 8 times the laminar value...'' 
 
Direct numerical simulations and experiments provide detailed data of the transition process that induces the increased skin friction, both in orderly and bypass breakdown to turbulence \citep{Kachanov1994}.
Orderly transition proceeds from upstream amplification of Tollmien–Schlichting instability waves, to secondary instability \citep{Herbert1988}, then breakdown of the elevated shear layers, and finally spreading of turbulence throughout the downstream boundary layer \citep{Sandham_kleiser_1992}.
Bypass transition, which is the focus of the present work, can take place at subcritical Reynolds numbers in presence of moderate levels of free-stream turbulence \citep{Alfredsson1994,Jacobs_durbin_2001}.
In this scenario, only low-frequency free-stream disturbances penetrate the boundary layer due to an effect known as shear-sheltering \citep{Hunt_Durbin_1999,Zaki_Saha_2009}.
The boundary-layer response comprises streamwise elongated energetic streaks, also termed Klebanoff modes \citep{Kendall1991}, whose amplification has been explained by vertical displacement of mean momentum \citep{Landahl1980}. The next stage is secondary instability of the streaky base state \citep{Andersson2001,Hack_zaki_2014} and breakdown into turbulent spots \citep{Brandt2004,Zaki2013}.
In both orderly and bypass transition processes, the final stage features spreading turbulent patches and an associated high wall stress.

Statistical approaches have been adopted to interpret the enhanced skin friction at transition onset.  
These methods often start from the Reynolds-averaged equations, and express the mean skin friction in terms of ensemble-averaged quantities, e.g.\, the laminar value and contribution due to the Reynolds stress.
By integrating the mean momentum equation in wall-normal direction, \cite{Fukagata2002} derived an expression of the mean skin friction for fully-developed turbulent flows.
They concluded that the dominant role is by the near-wall Reynolds shear stress which is related to vortical structures in the wall layer.
That work was further refined by \cite{Johnson2019} to differentiate the laminar and turbulent contributions in a developing boundary layer. 
A different decomposition can be obtained by integrating the mean vorticity equations \citep{Yoon2016}. 
In doing so, the mean skin friction is related to the advection of vorticity, vortex stretching, and the viscous diffusion effect.
Ultimately, these quantitative approaches express the mean skin friction in terms of a balance equation involving statistical flow quantities. 
The time-dependent dynamics, or flow history, is not exposed and as a result a causal relation between skin friction and individual flow events is not evident in these expressions.

Another perspective on drag increase at transition onset is provided by detailed visualizations of the instantaneous flow structures and the history of their evolution. Such visualizations can be quantitatively analysed by conditionally sampling events or computing correlations of interest \citep{Nolan_Zaki_2013,Marxen_Zaki_2019}. 
Through these approaches, various candidate mechanisms for the enhanced skin friction can be posited:
(i) Since the mean flow accelerates along the streamwise direction during transition, there must be an average wall-normal motion that brings high-momentum fluid towards the wall.
(ii) As turbulent spots impinge onto the wall, they transport turbulent vorticity towards the wall and could lead to a higher instantaneous skin friction. 
(iii) Due to the balance between pressure gradient and wall vorticity flux, the instantaneous pressure gradient may drive the increase of wall stress.
This conjecture is supported by the observation that the wall stress maximum is accompanied by strong pressure gradient in turbulent boundary layers \citep{Andreopoulos_Agui_1996,Ghaemi_scarano_2013}. 
(iv) The mechanism of enhanced skin friction during transition may be analogous to the vigorous sweep events in turbulent flows, which are associated with quasi-streamwise vorticies and the autonomous cycle in the near-wall layer \citep{Jimenez2007,Sheng2009}. 
These interpretations intuitively relate the enhanced skin friction to other flow events, but lack a definitive quantitative connection. 

\begin{figure}
	\centering
	\includegraphics[width=\textwidth]{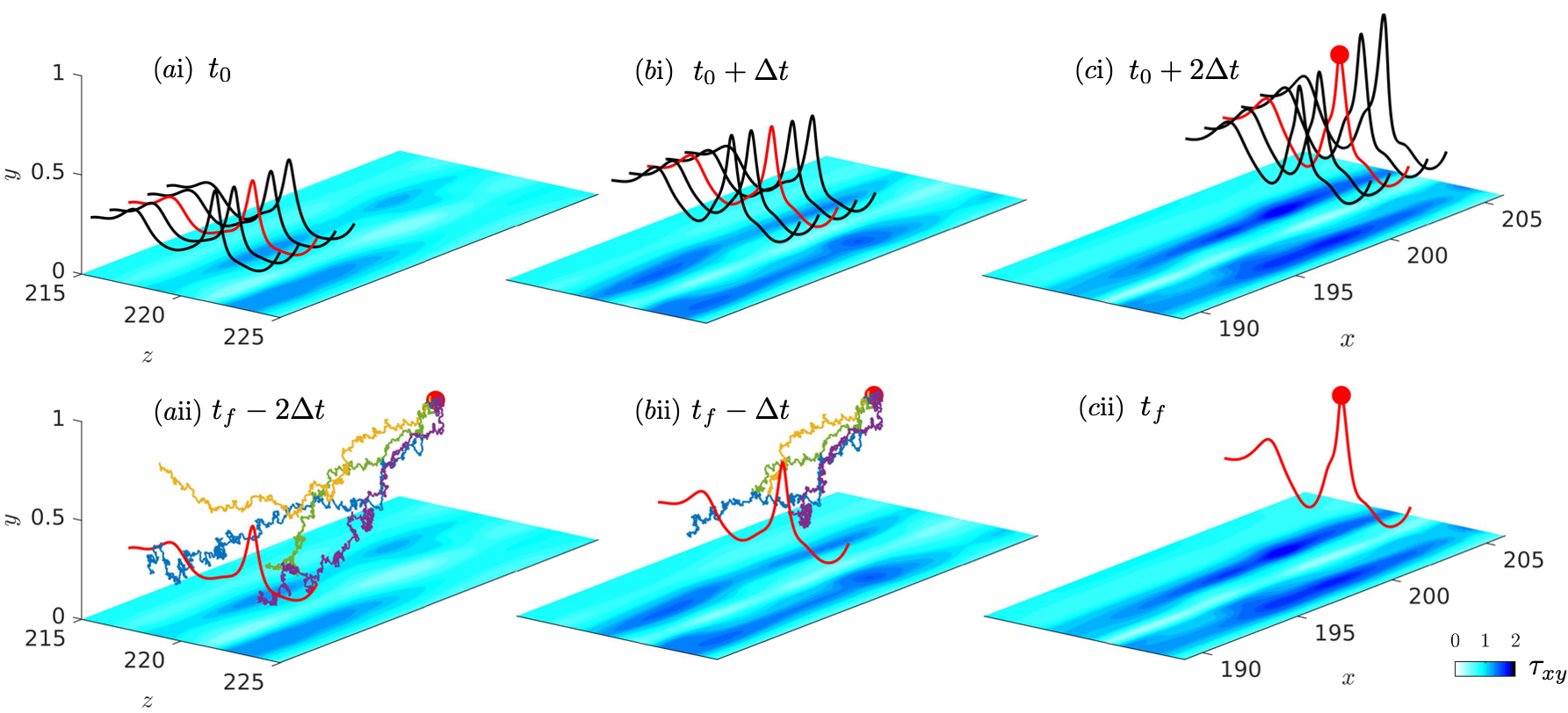}
	\caption{($a$i-$c$i) Forward evolution of vortex lines. ($c$ii-$a$ii) Backward evolution of stochastic Lagrangian trajectories that determine the origin of vorticity vector marked by the red dot in ($c$ii). Contours show the wall stress $\tau_{xy}$.
	}
	\label{fig:vort_evolution}
\end{figure}

In the present work we will adopt a precise quantitative approach that relates the enhanced wall stress to the preceding flow dynamics, or flow history. To motivate our approach, we consider a sample visualization of wall-bounded flow in figure \ref{fig:vort_evolution}i: Vortex lines are visualized above a pair of stress maxima in a transitional boundary layer from time $t_0$ to $t_f = t_0 + 2\Delta t$ (panels $a$i-$c$i). 
The vortex lines appear to be advected downstream and stretched in the wall-normal direction; and it may be tempting to relate the vorticity at the identified red point (panel $c$i) to the tips of the  earlier red vortex lines.  Such interpretation is, however, valid only in inviscid flows.
Due to viscosity, vortex lines are not material lines. 
Instead of being advected and stretched only, the vortex lines in panel $a$i also diffuse and affect the entire vorticity field at later times.
Conversely, the vorticity at the marked location in panel $c$i is affected by earlier vorticity at various locations that all contribute due to diffusion.

An exact approach to vorticity dynamics in viscous incompressible fluids has been provided in the recent mathematical work by \cite{Constantin2008,Constantin2011}, who expressed the vorticity as the expectation of a stochastic Cauchy invariant, evaluated along stochastic Lagrangian trajectories evolved backward in time.
Their formulation is derived by assuming a modest smoothness of the Navier-Stokes solutions (velocity twice-differentiable in space).
The stochastic Lagrangian approach was implemented numerically and validated in turbulent channel flow \citep{Eyink2020_theory}, and subsequently applied to analyze the origin of vorticity in ``sweep'' and ``ejection'' events
\citep{Eyink2020_channel}.
To study skin friction in this framework, the wall stress is expressed in terms of the wall vorticity,  $\btau_W=\nu\bomega_W\times\hat{\boldsymbol n},$ where $\hat{\boldsymbol n}$ is the inward-pointing normal at the wall. 
Unfortunately, the original formulation by \cite{Constantin2011} employs Dirichlet boundary conditions for the vorticity and is thus inadequate to describe the origin of the wall vorticity itself. 
Therefore, we augment the stochastic Lagrangian analysis with the Neumann boundary condition determined by the local vorticity flux \citep{Lighthill}, and exploit it to determine the origin of enhanced skin friction in the transitional boundary layer.

To aid intuition, a sample illustration of the stochastic Lagrangian approach is provided in figure \ref{fig:vort_evolution}ii. 
In order to discover the origin of the vorticity marked by the red dot in panel $c$ii at time $t_f$, particles are released from this point.  Their stochastic Lagrangian evolution in traced in backward time (panel $b$ii and $a$ii), where the stochastic diffusion represents exactly the action of viscosity.  The particles spread over space rather than land onto a single point on a vortex line earlier at earlier times.  
Therefore, the vorticity at time $t_f$ depends on the vorticity at many locations sampled by the particles at time $t_0$, unlike the perhaps appealing but inaccurate visual interpretation of material vortex lines in figure \ref{fig:vort_evolution}i. The stochastic Cauchy analysis provides a precise quantitative account of how the vorticity at earlier times and positions contribute to the final value, and the dominant mechanism that generates the target vorticity can be extracted.

The content of this paper is outlined as follows.
In \S\ref{sec:CauchyReview}, we provide a detailed explanation of the Cauchy invariants for inviscid and viscous fluid.
The stochastic Lagrangian analysis with Neumann boundary condition is elaborated in \S\ref{sec:SCauchy}, and the numerical procedures to compute the stochastic Cauchy invariant are summarized.
In \S\ref{sec:setup}, we introduce the direct numerical simulation (DNS) dataset of bypass transition, followed by an explanation about how the high wall-stress events of interest are selected.
The results obtained from the stochastic Cauchy analysis are presented in \S\ref{sec:results}.
We first focus on a particular event of suddenly increased skin friction, visualize the locations of Lagrangian particles, and evaluate the statistics of different terms in the stochastic Cauchy invariant.
Then a quantitative analysis across numerous similar events is performed.
The main conclusions drawn from our analysis are summarized in \S\ref{sec:results}.

\section{Methodology}
\label{sec:method}

Considering a location and time of interest $(\boldsymbol x,t)$, our objective is to explore the origin of the corresponding vorticity $\boldsymbol \omega(\boldsymbol x,t)$.  The particular point of interest here is at the wall, where the vorticity is proportional to the shear stress at the no-slip boundary.

\subsection{Cauchy invariants for incompressible Euler and Navier-Stokes equations}
\label{sec:CauchyReview}

Our study makes use of the Lagrangian formulation of vorticity dynamics by \cite{cauchy1815theorie}. This is mathematically 
equivalent to the circulation theorem of \cite{kelvin1868vi} but, rather than describing the evolution of surface integrals 
of vorticity, it applies to individual vorticity vectors and expresses their remarkable ``frozen-in'' properties for ideal 
incompressible Euler flows. More recently, these invariants have been extended to viscous Navier-Stokes flows 
\citep{Eyink2020_theory} based on a stochastic Lagrangian formulation of the incompressible Navier-Stokes equations  
\citep{Constantin2008,Constantin2011} and applied to analyze the viscous vorticity dynamics of near-wall sweep and ejection 
events in turbulent channel flow \citep{Eyink2020_channel}. 
We begin here with a basic  
introduction to these invariants for Euler equations and their stochastic extension to 
Navier-Stokes.

We start with the classical theory for an inviscid fluid described by the incompressible Euler equations 
and without boundaries. Consider the Lagrangian trajectory $\bX(\ba,t)$ of a 
fluid particle whose initial position is $\ba$ at time zero, which satisfies
\begin{equation} \frac{d\bX(\ba,t)}{dt}=
\bu(\bX(\ba,t),t), \quad \bX(\ba,0)=\ba.
\label{Xeq} \end{equation}
It is a straightforward consequence of the Helmholtz equation that vorticity satisfies 
 \begin{equation} 
 \frac{d}{dt} {\boldsymbol \omega}(\bX(\ba,t),t)=
{\boldsymbol \omega}(\bX(\ba,t),t)\cdot\nabla_{\boldmath x}\bu(\bX(\ba,t),t). 
\label{Lag-omega-eq} \end{equation}
This result has the intuitive meaning that vorticity vectors are transported by the ideal flow exactly as infinitesimal 
material line vectors ${\boldsymbol l}(t)$ with initial vectors ${\boldsymbol l}_0$ located at point $\ba$ at time zero:  
\begin{equation} 
{\boldsymbol l}(t)=\bX(\ba+{\boldsymbol l}_0,t) 
-\bX(\ba,t)\simeq {\boldsymbol l}_0\cdot \nabla_{\boldmath a}\bX(\ba,t).
\end{equation} 
That ${\boldsymbol l}(t)$ satisfies the same evolution equation as ${\boldsymbol \omega}(\bX(\ba,t),t)$
follows from the result 
\begin{equation} \frac{d}{dt}\nabla_{\boldmath a}\bX(\ba,t)=\nabla_{\boldmath a}\bX(\ba,t) \cdot
\nabla_{\boldsymbol x}\bu(\bX(\ba,t),t)
\end{equation}
obtained by applying the gradient $\nabla_{\boldmath a}$ to equation \eqref{Xeq}. 
This observation allows the Lagrangian 
evolution equation \eqref{Lag-omega-eq} to be exactly integrated as 
\begin{equation} 
{\boldsymbol \omega}(\bX(\ba,t),t)
={\boldsymbol \omega}(\ba,0)\cdot \nabla_{\boldmath a}\bX(\ba,t), 
\label{Cauchy-form} 
\end{equation}
which is the so-called {\it Cauchy formula} for vorticity, originally derived by \cite{cauchy1815theorie} in the Lagrangian 
formulation of the Euler fluid without using the Helmholtz equation. It follows immediately that the initial vorticity 
${\boldsymbol \omega}(\ba,0)$ can be written as 
\begin{equation} 
{\boldsymbol \omega}(\ba,0)
=(\nabla_{\boldmath a}\bX(\ba,t))^{-\top} \cdot {\boldsymbol \omega}(\bX(\ba,t),t)
\label{Cauchy-inv} 
\end{equation}
where the superscript $(\bullet)^{-\top}$ represents inverse transpose of a matrix. The initial vorticity is thus expressed 
as a formally conserved quantity of the Lagrangian flow $\bX(\ba,t),$ the so-called {\it Cauchy invariant}, satisfying 
$(d/dt)\omega(\ba,0)=0$ for each position label $\ba$.
For the interested reader, we have included a brief discussion of the connection to geometric fluid mechanics in Appendix \ref{sec:geometric}.

\begin{figure}
	\centering
	\includegraphics[width=\textwidth]{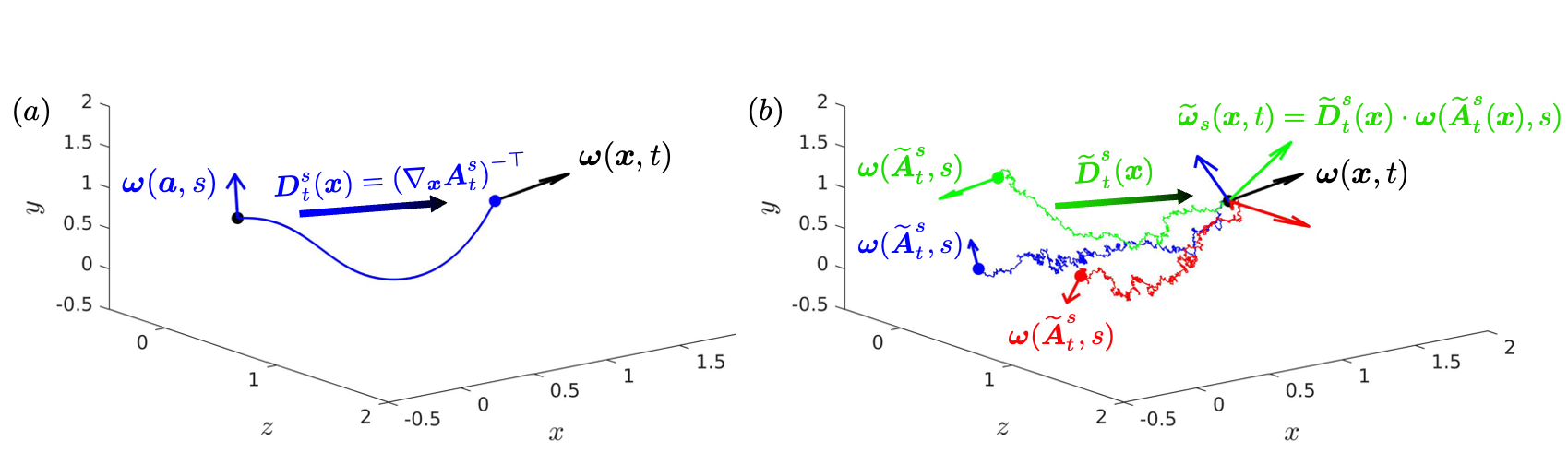}
	\caption{($a$) Lagrangian trajectory in inviscid fluid. The vorticity $\boldsymbol \omega(\boldsymbol x,t)$ originates from an earlier vorticity vector $\boldsymbol \omega(\boldsymbol a,s)$ through stretching and tilting. ($b$) Stochastic Lagrangian trajectories in viscous fluid. Earlier vorticity vectors $\boldsymbol \omega (\widetilde{\boldsymbol A}_t^s,s)$ sampled by the stochastic trajectories are transported to $\boldsymbol \omega_s (\boldsymbol x,t)$ and ensemble-averaged to obtain the target vorticity $\boldsymbol \omega(\boldsymbol x,t)$.
	}
	\label{fig:trajectory}
\end{figure}

Because of the time-reversibility of the Euler fluid equations, the Cauchy formula \eqref{Cauchy-form} and the Cauchy 
invariants \eqref{Cauchy-inv} are valid also backward in time, providing an exact reconstruction of the vorticity 
$\bomega(\bx,t)$ from its value $\bomega(\ba,s)$ at an earlier time $s<t.$
Here it is useful to generalize the prior discussion and introduce Lagrangian particle positions $\bX^s_t(\ba)$ at time $t$ which are ``labelled'' 
by positions $\ba$ at time $s$ rather than at time 0. 
The superscript thus refers to the labelling time, and the subscript denotes the specific time when the particle location is considered.
These particles positions evolve according to the same equation \eqref{Xeq} but now satisfy
$\bX^s_s(\ba)=\ba.$
By introducing the ``back-to-labels'' map $\bx\mapsto\ba=\bA^s_t(\bx)$,
which is inverse to $\ba\mapsto\bx=\bX^s_t(\ba)$, the Cauchy formula \eqref{Cauchy-form} can then be expressed as
\begin{equation}
    \label{eq:inviscid_Cauchy}
    \boldsymbol \omega(\boldsymbol x,t) = \boldsymbol D_t^s(\boldsymbol x) \cdot \boldsymbol \omega(\boldsymbol A_t^s(\boldsymbol x),s), \quad \boldsymbol D_t^s(\boldsymbol x) = (\nabla_{\boldsymbol a}\bX^s_t)^\top = \left(\nabla_{\boldsymbol x} \boldsymbol A_t^s \right)^{-\top}, 
\end{equation}
where $\boldsymbol D_t^s$ is the so-called ``deformation matrix'' quantifying the vorticity stretching and tilting from $s$ to $t.$
It is furthermore useful to observe that $\bA^s_t=\bX^t_s,$ since evolving the particle backward in time from $t$ to an earlier time $s$ via the flow $\bX^t_s$
exactly recovers its label $\ba$ at time $s.$ Thus, the back-to-labels map $\bA^s_t$ satisfies the same advection equation \eqref{Xeq} 
as does $\bX^s_t(\ba),$ but now in the time variable $s$:
\begin{equation}
    \label{eq:inviscid_dAds}
    \frac{d \boldsymbol A_t^s(\boldsymbol x)}{ds} = \boldsymbol u\left(\boldsymbol A_t^s(\boldsymbol{x}),s\right), \quad s < t; \quad \boldsymbol A_t^t(\boldsymbol x) = \boldsymbol x. 
\end{equation}
Because $\bA^s_t=\bX^t_s,$ the formula (\ref{eq:inviscid_Cauchy}) formally expresses the vorticity $\bomega(\bx,t)$ as a Cauchy invariant of the backward-in-time evolution,
independent of the choice of $s<t.$ 
This interpretation of (\ref{eq:inviscid_Cauchy}) is shown schematically in figure \ref{fig:trajectory}$a$, where $\boldsymbol \omega(\boldsymbol x,t)$ is invariant along the backward Lagrangian trajectory (blue curve): no matter the stretching or tilting along the path or the choice of initial time $s$, the vorticity $\boldsymbol \omega(\boldsymbol x,t)$ is always equal to $\left(\nabla_{\boldsymbol x} \boldsymbol A_t^s \right)^{-\top} \boldsymbol \omega(\boldsymbol A_t^s(\boldsymbol x) ,s)$.

Physical fluids always involve viscosity, however, so that the remarkable Lagrangian properties enjoyed by vorticity in inviscid flows do not seemingly apply to most real-world flows, except approximately in flow regimes and space-time regions where viscous effects are negligible. Recently, it was pointed out by \cite{Eyink2020_theory} that the Cauchy invariant (\ref{eq:inviscid_Cauchy}) holds in a probabilistic sense for viscous incompressible fluids, based on a stochastic representation of incompressible Navier-Stokes solutions derived by \cite{Constantin2008}.
In this formulation, viscous diffusion and nonlinear advection of vorticity are simultaneously represented by the stochastic Lagrangian particle trajectories $\widetilde \bA_t^s(\boldsymbol x)$, which satisfy backward in time a Langevin equation with noise term proportional to the square-root of viscosity:
\begin{equation}
    \label{eq:dAds_nowall}
    \hat{\mathrm{d}} \widetilde{\boldsymbol{A}}_{t}^{s}(\boldsymbol{x})=\boldsymbol{u}\left(\widetilde{\boldsymbol{A}}_{t}^{s}(\boldsymbol{x}), s\right) \mathrm{d} s+\sqrt{2 \nu} \hat{\mathrm{d}} \widetilde{\boldsymbol{W}}(s), \quad s < t; \quad \widetilde{\boldsymbol{A}}_{t}^{t}(\boldsymbol{x}) = \boldsymbol x.
\end{equation}
Here $\boldsymbol u$ is any solution to the incompressible Navier-Stokes equations, $\nu$ is the kinematic viscosity, $\hat {\mathrm d}$ denotes the backward It$\bar{{\rm o}}$ differential, and $\widetilde{\boldsymbol{W}}(s)$ is a vector Brownian motion.
Note that the backward It$\bar{{\rm o}}$ differential is just the time-reverse 
of the more widely known forward It$\bar{{\rm o}}$ differential (see \cite{Constantin2011},
section 4).
Samples of backward-in-time stochastic Lagrangian trajectories starting from $(\boldsymbol x,t)$ are shown in figure \ref{fig:trajectory}$b$.
Due to stochasticity, the Lagrangian particles significantly diverge once leaving the starting location $\boldsymbol x$, and the trajectories are not smooth, especially if compared against figure \ref{fig:trajectory}$a$.
\cite{Constantin2008} proved that an expectation
$\mathbb{E}$ over the ensemble of Brownian motions yields the solution $\bomega(\bx,t)$ of the viscous Helmholtz equation as
\begin{equation}
    \label{eq:SCauchy_nowall}
    \boldsymbol \omega(\boldsymbol x,t) = \mathbb{E}\left[ \widetilde\bD_t^s(\boldsymbol x) \cdot \boldsymbol \omega(\widetilde\bA_t^s(\boldsymbol x),s) \right] \vcentcolon= \mathbb{E}\left[\widetilde{\boldsymbol \omega}_s(\bx,t) \right]. 
\end{equation}
where $\widetilde{\bD}_t^s(\boldsymbol x) = \left(\nabla_{\boldsymbol x} \widetilde{\bA}_t^s \right)^{-\top}$.
The interpretation of equation \eqref{eq:SCauchy_nowall} is provided in figure \ref{fig:trajectory}$b$. 
Starting from $(\boldsymbol x,t)$, the stochastic Lagrangian trajectories (colored curves) are integrated backward in time until $s$. 
The vorticity vectors sampled by the Lagrangian particles (colored arrows annotated with $\boldsymbol \omega(\widetilde\bA_t^s(\boldsymbol x),s)$) constitute the origin of the target vorticity $\boldsymbol \omega(\bx,t)$ (black arrow). 
Quantitatively, the earlier vorticities are transported by the deformation matrix $\widetilde{\bD}_t^s$ to obtain the stochastic Cauchy invariant $\widetilde{\boldsymbol \omega}_s(\bx,t)$, whose ensemble average is equivalent to the vorticity of interest $\boldsymbol \omega(\bx,t)$.
Since the Navier-Stokes equations are time-irreversible, 
it is natural that such stochastic invariants exist only backward in time and the formula \eqref{eq:SCauchy_nowall}
yields a causal representation of the vorticity $\bomega(\bx,t)$ in terms of its values $\bomega(\ba,s)$ at 
each earlier time $s<t.$ This formula thus represents exactly how vortex lines move, or more precisely evolve, 
in a viscous fluid, through the  combination of both nonlinear advection and viscous diffusion.

In wall-bounded flows, the treatment of the stochastic trajectories when they reach the wall is equivalent to the choice of the boundary conditions for the viscous Helmholtz equations.
\cite{Constantin2011} proved that stopping the trajectories at the wall is equivalent to the Dirichlet boundary condition of the vorticity. 
Unfortunately, this formulation is inadequate for the purposes of our current investigation, because the problem is precisely 
to understand the evolution of the wall stress $\btau_W=\nu\bomega_W\times\bn$ and thus $\bomega_W$
cannot be taken as given. In this context it is appropriate instead to adopt Neumann boundary conditions, where 
the wall vorticity-flux is prescribed. The derivation of a stochastic Lagrangian 
representation for the Navier-Stokes vorticity with such boundary conditions is one of the main results of the present 
paper and is accomplished in the following section.

\subsection{Stochastic Cauchy invariant with Neumann boundary condition}
\label{sec:SCauchy}

\begin{figure}
	\centering
	\includegraphics[width=0.8\textwidth]{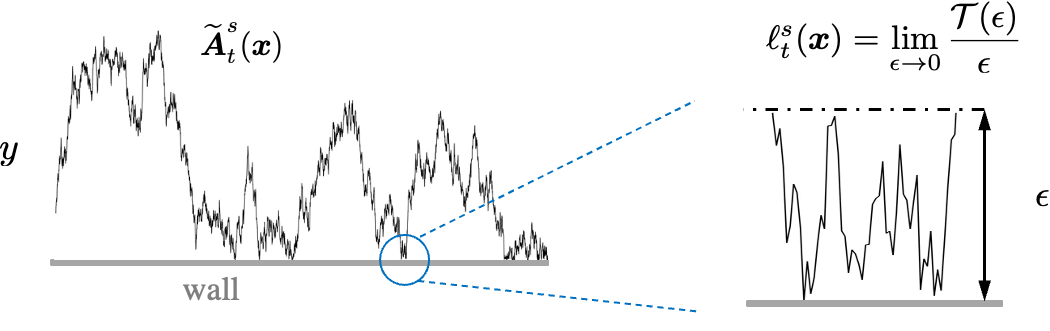}
	\caption{Schematic of the stochastic Lagrangian trajectory with Neumann boundary condition, and the physical interpretation of the boundary local time density. Given a distance from the wall $\epsilon$, $\mathcal T(\epsilon)$ denotes the time spent by the particle within $\epsilon$.
	}
	\label{fig:localtime}
\end{figure}

Here we adopt a Neumann boundary condition determined by vorticity source density at the wall, which is equivalent to reflecting the stochastic Lagrangian trajectories every time when they reach the wall \citep{Drivas2017}.
At no-slip boundaries, most terms in the momentum equation vanish, except the pressure gradient and the viscous diffusion terms.
Therefore, in the case of a flat wall considered here, the diffusion of vorticity, or so-called Lighthill source 
\citep{Lighthill,Panton2006}, has tangential components balanced by the pressure-gradient 
\begin{equation}
    \label{eq:Lighthill}
    \boldsymbol \sigma \vcentcolon=-\nu \hat{\boldsymbol n} \cdot \nabla \boldsymbol{\omega}|_{\mathrm w}= - \hat {\boldsymbol n} \times \nabla p |_{\mathrm w}.
\end{equation}
The stochastic trajectory (\ref{eq:dAds_nowall}) is augmented with a reflecting boundary condition,
\begin{equation}
    \label{eq:dAds}
    \hat{\mathrm{d}} \widetilde{\boldsymbol{A}}_{t}^{s}(\boldsymbol{x})=\boldsymbol{u}\left(\widetilde{\boldsymbol{A}}_{t}^{s}(\boldsymbol{x}), s\right) \mathrm{d} s+\sqrt{2 \nu} \hat{\mathrm{d}} \widetilde{\boldsymbol{W}}(s) - \nu \hat{\boldsymbol n}\left(\widetilde{\boldsymbol{A}}_{t}^{s}(\boldsymbol{x}), s\right) \hat{\mathrm d} \ell_{t}^{s}(\boldsymbol{x}), \quad s < t,
\end{equation}
enforced by the backward boundary-local time density $\ell_t^s(\boldsymbol x)$, 
formally defined as
\begin{equation}
    \label{eq:localtime_l}
    \ell_{t}^{s}(\boldsymbol{x})=\int_{t}^{s} d r \int_{\mathrm w} d S(\mathbf{z})\ \delta^{3}\left(\mathbf{z}-\widetilde\bA_{t}^{s}(\boldsymbol{x})\right), \quad s<t
\end{equation}
where $\delta(\bullet)$ is the Dirac delta function, and $\int_{\mathrm w} dS(z)$ represents surface integration over the wall \citep{Lions_Sznitman_1984,Burdzy2004,Drivas2017}.
By definition, $\ell_t^s(\boldsymbol x)$ has the dimension of time divided by length. 
The physical interpretation of $\ell_t^s(\boldsymbol x)$ is provided in figure \ref{fig:localtime}. 
A reflected stochastic Lagrangian trajectory governed by (\ref{eq:dAds}) is shown on the left. Given a distance from the wall $\epsilon$, the backward time that the particle spent within $y \in [0,\epsilon]$ is $\mathcal T(\epsilon)$ (defined to be negative).
Due to stochasticity of the trajectory, as $\epsilon \rightarrow 0$, the ratio $\mathcal T(\epsilon) / \epsilon$ converges to a finite value, the boundary local time density $\ell_t^s(\boldsymbol x)$.
Therefore, $\ell_t^s(\boldsymbol x)$ remains non-positive and unchanged when the particle does not reach the wall.
If reflection occurs, $\ell_t^s(\boldsymbol x)$ decreases, and the last term in equation (\ref{eq:dAds}) quantifies the reflected distance from the wall. 

We prove in appendix \ref{sec:proof} that the expectation of the stochastic Cauchy invariant supplemented with a term from the Lighthill source is conserved in backward time $s<t$,
\begin{equation}
    \label{eq:SCauchy}
    \boldsymbol \omega(\boldsymbol x,t) = \mathbb{E} \left[\widetilde {\boldsymbol \omega}_s(\boldsymbol x,t) \right] = \mathbb{E} \left[ \widetilde \bD_{t}^{s}(\boldsymbol{x}) \cdot  \boldsymbol{\omega}\left(\widetilde \bA_{t}^{s}(\boldsymbol{x}), s\right)+ \widetilde {\boldsymbol L}_t^s(\boldsymbol x) \right],
\end{equation}
where matrix $\widetilde \bD_{t}^{s}(\boldsymbol{x})$ is given by the solution of the final-value problem
\begin{equation}
\frac{d}{ds} \widetilde \bD_{t}^{s}(\boldsymbol{x}) = -\widetilde \bD_{t}^{s}(\boldsymbol{x})
(\boldsymbol \nabla_x \bu(\widetilde \bA_{t}^{s}(\boldsymbol{x}),s))^\top  
\qquad \widetilde \bD_{t}^{t}(\boldsymbol{x})={\boldsymbol I},  
\label{Deq} \end{equation}  
and where the source term is 
\begin{equation}
    \label{eq:SCauchy_L}
    \widetilde{\boldsymbol L}_t^s(\boldsymbol x) = \int_{s}^{t} \widetilde \bD_{t}^{r}(\boldsymbol{x}) \cdot  \boldsymbol{\sigma}\left(\widetilde \bA_{t}^{r}(\boldsymbol{x}), r\right) \hat{\mathrm d} \ell_{t}^{r}(\boldsymbol{x}).
\end{equation}
The two terms in (\ref{eq:SCauchy}) quantify the contributions to $\boldsymbol \omega(\boldsymbol x,t)$ of interior vorticity deformation and of the wall vorticity flux, respectively.
Thus, the Lighthill source \eqref{eq:Lighthill} is sampled every time the particle hits the boundary and is reflected.

Although we focus in this work on a developing boundary layer over a flat wall, our method of solving 
stochastic differential equations with reflecting boundary conditions applies to flow domains with curved walls. 
Details can be found in the paper of \cite{Lions_Sznitman_1984}, which treated any domain $\Omega$ 
whose boundary is a smooth manifold. In fact, their analysis covered a more general class of ``admissible'' 
open domains $\Omega$ which includes those whose boundary is piecewise smooth with components intersecting 
at convex interior angles. The construction of \cite{Lions_Sznitman_1984} established globally in time both the existence and uniqueness of the reflected diffusion process and the boundary local-time density for all such domains.
Their result already covers a large number of flows of physical interest, including those in which the wall is mathematically smooth but ``hydraulically rough'', e.g. when the height of the wall is given by a sinusoidal profile.
Our analysis thus carries over to a large class of flow domains.

The stochastic Cauchy invariant (\ref{eq:SCauchy}) can be numerically evaluated using a Monte-Carlo scheme and Euler-Maruyama method for time discretization.
At discrete times $s = s_k \coloneqq t - k(\Delta s), k=1,2,3,...$, the particle locations $\widetilde \bA_t^{s_k}$, the deformation matrix $\widetilde \bD_t^{s_k}$, and the wall contribution $\widetilde {\boldsymbol L}_t^{s_k}$ are obtained through backward integration,
\begin{eqnarray}
    \label{eq:disc_A}
    \widetilde{\boldsymbol{A}}_{t}^{s_{k}}(\boldsymbol{x}) &=& \widetilde{\boldsymbol{A}}_{t}^{s_{k-1}}(\boldsymbol{x})-\boldsymbol{u}\left(\widetilde{\boldsymbol{A}}_{t}^{s_{k-1}}(\boldsymbol{x}), s_{k-1}\right) \Delta s+\sqrt{2 \nu \Delta s} \widetilde{\boldsymbol {N}}_{k} - \nu \Delta \ell_{k} \hat{\boldsymbol y}, \\
    \label{eq:disc_D}
    \widetilde{\boldsymbol{D}}_{t}^{s_{k}} &=& \widetilde{\boldsymbol{D}}_{t}^{s_{k-1}} \cdot\left[\boldsymbol{I}+\left.\left(\boldsymbol{\nabla}_{x} \boldsymbol{u}\right)^{\top}\right|_{\left(\widetilde{\boldsymbol{A}}_{t}^{s_{k-1}}, s_{k-1}\right)} \Delta s\right], \\
    \label{eq:disc_L}
    \widetilde{\boldsymbol{L}}_{t}^{s_{k}} &=& \widetilde{\boldsymbol{L}}_{t}^{s_{k-1}}-\widetilde{\boldsymbol{D}}_{t}^{s_{k-1}} \cdot  \boldsymbol{\sigma}\left(\widetilde{\boldsymbol{A}}_{t}^{s_{k}}, s_{k-1}\right) \Delta \ell_{k}.
\end{eqnarray}
In equation (\ref{eq:disc_A}), $\widetilde{\boldsymbol N}_k$ is a three-dimensional normal random vector with mean zero and covariance matrix $\boldsymbol I$, independently sampled for each step $k=1,2,3...$
The increment of the boundary local time density is denoted as $\Delta \ell_k = \ell_t^{s_k} - \ell_t^{s_{k-1}}$ and details about its evaluation are provided in Appendix \ref{sec:localtime}.
Given a fully-resolved Navier-Stokes solution $\boldsymbol u(\boldsymbol x,t)$, equations (\ref{eq:disc_A}-\ref{eq:disc_L}) can be exploited to evaluate the stochastic Cauchy invariant (\ref{eq:SCauchy}) in backward time.
The numerical procedures are summarized in Algorithm \ref{alg:SCauchy}.
In all the examined cases, the number of particles employed for Monte Carlo evaluation of the expectation is always $N_p=10^4$.

\begin{algorithm}[h]
	\SetAlgoLined
	\textbf{Step 1}: Initialization\;
	\Indp
	\textbullet~Set the number of Lagrangian particles $N_p$ \;
	\textbullet~For all the particles, set $\widetilde \bA_t^t = \boldsymbol x$, $\widetilde \bD_t^t = \boldsymbol I$, and $\widetilde {\boldsymbol L}_t^t = \boldsymbol 0$\;
	\Indm
	\textbf{Step 2}: Wall contribution\;
	\Indp
	\textbullet~Given $\widetilde \bA_t^{s_{k-1}}$ for each particle, evaluate $\nu \Delta \ell_k$, and skip this step if $\nu \Delta \ell_k = 0$\;
	\textbullet~If $\nu \Delta \ell_k < 0$ (particle is reflected from the wall), compute the Lighthill source and update the wall contribution (\ref{eq:disc_L})\;
	\Indm
	\textbf{Step 3}: Interior contribution\;
	\Indp
	\textbullet~Update the deformation matrix (\ref{eq:disc_D}) and particle location (\ref{eq:disc_A}) at time $s_k$ \;
	\textbullet~Evaluate the vorticity vector at location $\widetilde \bA_t^{s_{k}}$\;
	\Indm
	\textbf{Step 4}: Stochastic Cauchy invariant\;
	\Indp
	\textbullet~Compute the expectation of the stochastic Cauchy invariant $\widetilde {\boldsymbol \omega}_s(\boldsymbol x,t)$ (\ref{eq:SCauchy}) over all the particles \;
	\textbullet~Repeat Steps 2-4 until the earliest time of interest is reached.
	\caption{Stochastic Cauchy analysis with Neumann boundary condition.}
	\label{alg:SCauchy}
\end{algorithm}

\begin{figure}
	\centering
	\includegraphics[width=0.9\textwidth]{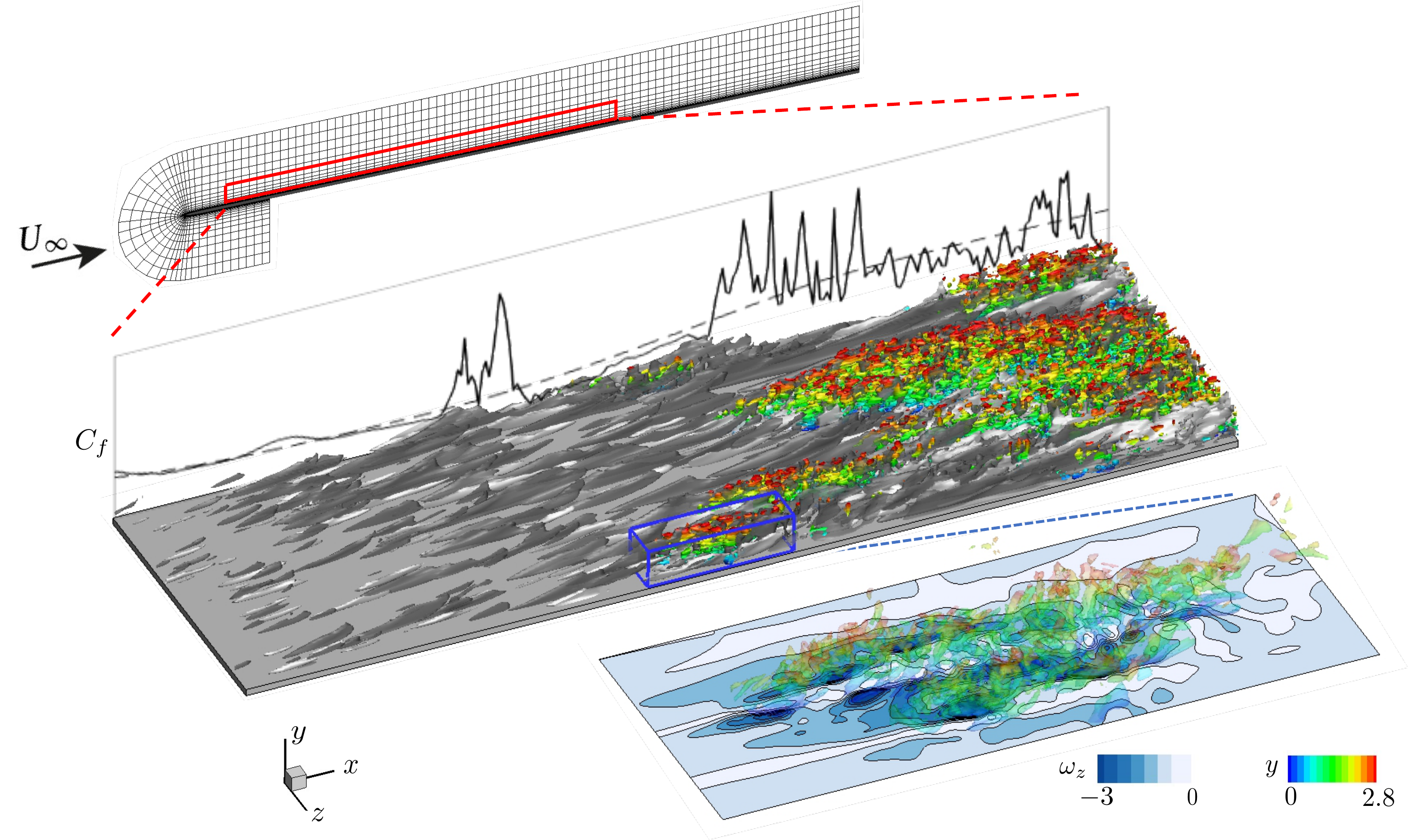}
	\caption{(Top) Computational grid for direct numerical simulation of transitional boundary layer. (Middle) Visualization of flow structure: high-speed ($u^{\prime} = 0.1$, white) and low-speed streaks ($u^{\prime} = -0.1$, dark gray); vortical structures identified using $\lambda_2$ criterion (red to green, $\lambda_2=-0.02$). Only half of the domain is shown in the spanwise direction. Mean (dashed line) and sample instantaneous (solid line) skin friction are shown on the side.
	(Bottom) Turbulent spot and the associated wall vorticity.
	}
	\label{fig:config}
\end{figure}

\subsection{Computational setup and events of interest}
\label{sec:setup}
Our study is performed using the transitional boundary layer dataset of the Johns Hopkins Turbulent Database (JHTDB) \citep{Wu2019}.
The computational domain and flow configuration are shown in figure \ref{fig:config}.
The dataset was produced from DNS of incompressible flow over a flat plate with an elliptical leading edge.
The streamwise, wall-normal, and spanwise coordinates are denoted by $x$, $y$ and $z$, and corresponding velocity components are $u$, $v$ and $w$.
The reference length-scale is half-thickness of the plate, $L$, and the reference
velocity is the incoming free-stream speed $U_{\infty}$.
At the upstream curved boundary of the domain, the inflow is a superposition of the
uniform velocity $U_{\infty}$ and homogeneous isotropic turbulence.
The free-stream turbulence decays as it
is advected towards the leading edge where its intensity reaches $Tu = 3\%$. 
The free-stream forcing of the boundary layer leads to the formation of amplifying streaks and sporadic breakdown into turbulent spots (figure \ref{fig:config}). As spots spread towards the wall and laterally, the wall friction in the footprint of the turbulence increases appreciably (solid line in the middle panel of figure \ref{fig:config} and the contours in the bottom panel). Accompanying the intermittent generation of turbulent spots, mean skin friction (dashed line) also increases to a turbulent level. 

\begin{table}
	\centering
	\begin{tabular}{c c}
	     DNS domain &  Analysis subdomain \\
	     \begin{tabular}{c c}
	         \hline
	         Domain Size  &  \ Grid points \\
	         \begin{tabular}{c}
			    \hline 
			    $(L_x,L_y,L_z)/L$ \\
			    (1050, 40, 240)
		     \end{tabular} & 
		     \begin{tabular}{c}
			    \hline
			    $(N_x,N_y,N_z)$ \\
			    (4097, 257, 2049)
		     \end{tabular} \\
	     \end{tabular} &
	     \begin{tabular}{c c}
	         \hline
	         Grid resolution  &  \ Flow statistics \\
	         \begin{tabular}{c c}
			    \hline
			    $(\Delta x,\Delta y_{\mathrm{min}},\Delta z)/L$ & $\Delta t U_{\infty}/L$ \\
			    (0.29, 0.0036, 0.12) & 0.25
		     \end{tabular} &
		     \begin{tabular}{c c}
			    \hline
			    $\delta_{99}$ & $Re_{\theta}$\\
			    2.7$-$7.4 & 209$-$650 
		     \end{tabular} \\
	     \end{tabular} \\
	\end{tabular}
	\caption{Domain size and the number of grid points for DNS within the curved domain. Grid resolution and flow statistics are reported for the transitional region $200 < x < 500$. $Re_{\theta} = U_{\infty} \theta/\nu$, where $\theta$ is the momentum thickness.}
	\label{table:setup}
\end{table}

To generate the database flow, the incompressible Navier-Stokes equations were solved on a curvilinear grid (see figure \ref{fig:config}) using a fractional-step method \citep{Rosenfeld1991}. 
A second-order volume-flux formulation was adopted for the spatial discretization.
The advection terms were treated explicitly by the Adams-Bashforth scheme, and the Crank-Nicolson scheme was adopted for the diffusion terms.
The pressure Poisson equation was solved using Fourier transform in the spanwise direction and multi-grid inversion for every spanwise wavenumber.
The algorithm has been applied in numerous studies of transitional and turbulent flows \citep{Zaki2013,Lee_Sung_Zaki_2017}.
The domain size and the number of grid points are summarized in Table \ref{table:setup}, and more details about the numerical method and flow statistics can be found at JHTDB.

In order to evaluate the stochastic Cauchy invariant (\ref{eq:SCauchy}, \ref{eq:disc_A}$-$\ref{eq:disc_L}) along stochastic Lagrangian trajectories, we adopt the JHTDB Web Service Interface to obtain the velocity and its derivatives at instantaneous particle locations.
The piecewise cubic Hermite interpolation (PCHIPInt) is adopted in time.
Since the DNS was performed using second-order methods, we choose the fourth-order Lagrangian interpolation (Lag4) for the \emph{getVelocity} subroutine, and the fourth-order finite-difference scheme (FD4Lag4) for \emph{getVelocityGradient} and \emph{getVelocityHessian} subroutines. 

\begin{figure}
	\centering
	\includegraphics[width=\textwidth]{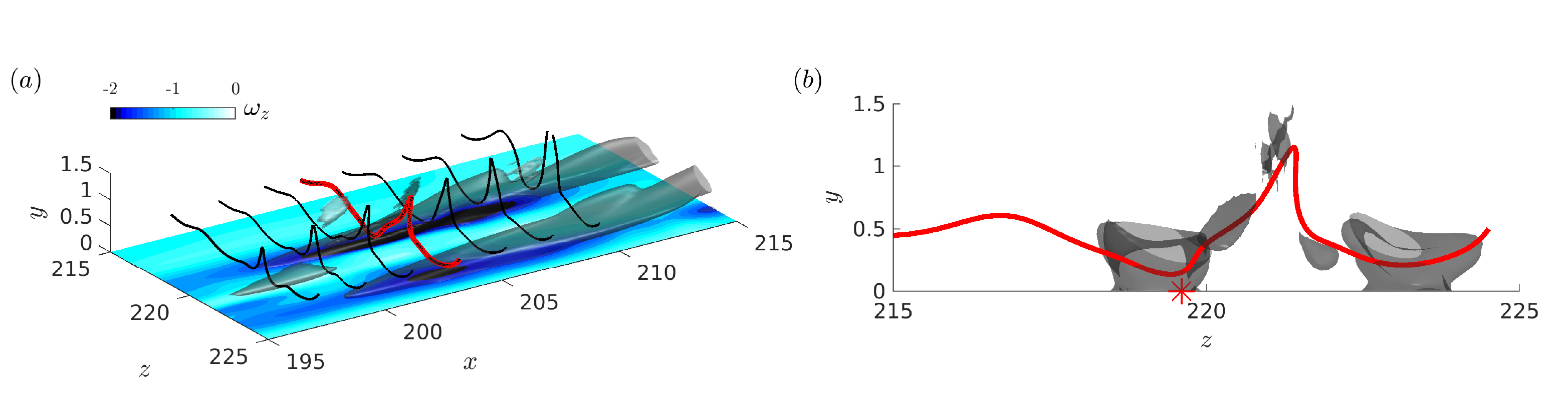}
	\caption{($a$) 3-D view and ($b$) end view of flow field around the stress maximum (marked by red asterisk in ($b$)).
	Black and red curves: vortex lines initiated at $y=0.5$;
	gray isosurface: $\omega_z = -1.5$;
	bottom plane: contour plot of spanwise wall vorticity.
	}
	\label{fig:event}
\end{figure}

\begin{figure}
	\centering
	\includegraphics[width=\textwidth]{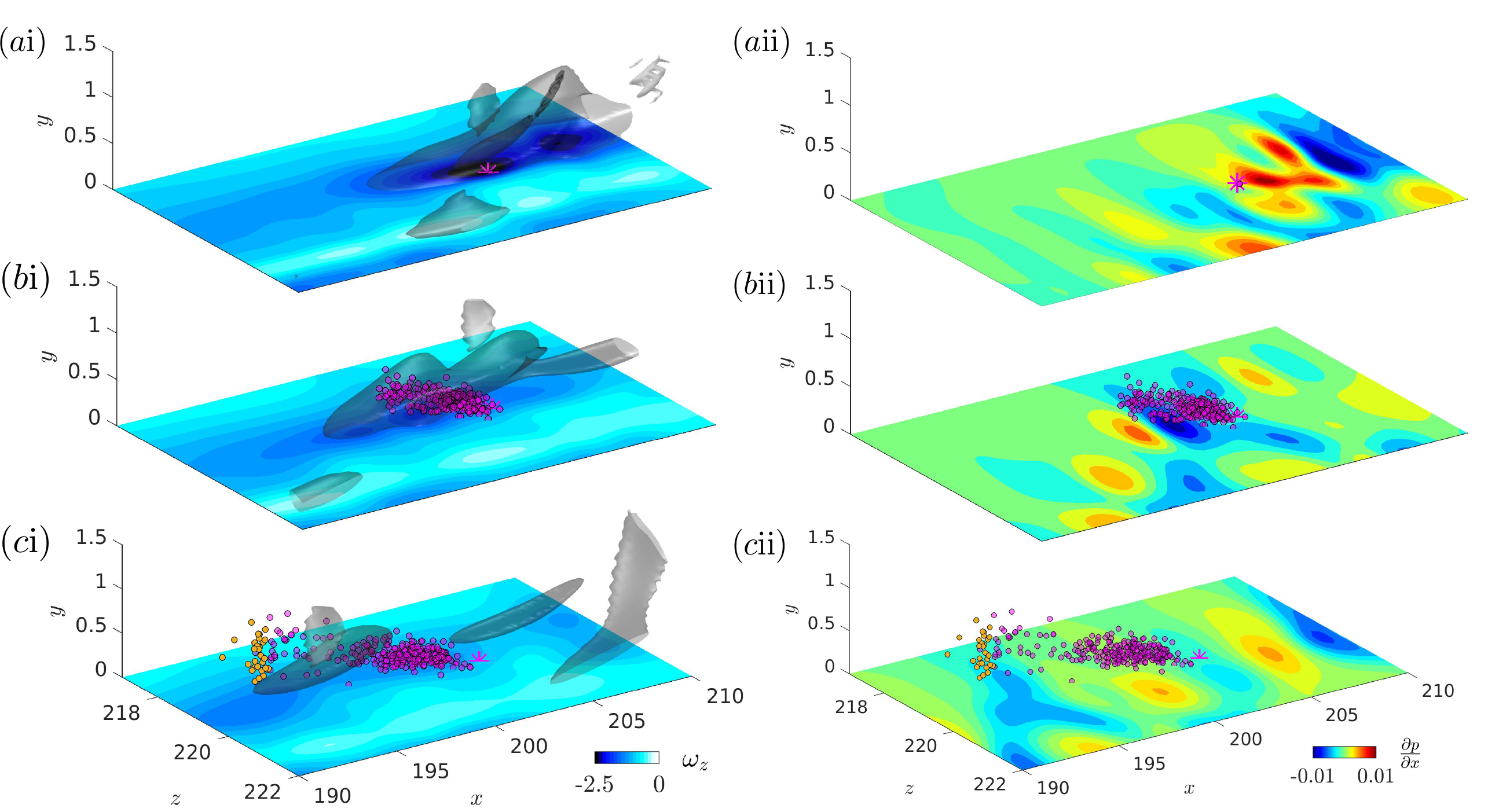}
	\caption{Instantaneous particle locations at ($a$-$c$) $\delta s = s - t =$ $0$, $-10$, $-20$. The bottom plane shows (i) spanwise wall vorticity and (ii) 
	spanwise vorticity source. 
	Gray isosurface: $\omega_z = -1.5$.
	Particles in ($c$) are separated into (magenta) near-wall cluster and (yellow) outer cluster}
	\label{fig:particles}
\end{figure}

In order to extract events that represent the enhanced skin friction, we search for local maxima of streamwise wall shear stress $\tau_{xy} = \nu \partial_y u = -\nu \omega_z$ within the transitional region $200 < x < 500$.
The grid resolution and flow statistics for the analysis subdomain are shown in Table \ref{table:setup}.
For each local maximum of wall stress, we keep track of its temporal evolution $\omega_m(t)$ and select the time when
\begin{equation}
    \label{eq:criteria}
    |\omega_{m}(t)| \leq \max_x \{ \left| \langle \omega_{z}\rangle \right|+ \omega^{\prime}_{z,\mathrm{rms}}  \} , \quad  \text{and} \quad
    |\omega_{m}(t+\Delta t)| > \max_x \{ \left| \langle \omega_{z}\rangle \right|+ \omega^{\prime}_{z,\mathrm{rms}} \}.
\end{equation}
Here $\langle \omega_{z} \rangle$ and $\omega^{\prime}_{z,\mathrm{rms}}$ are the mean and root-mean-square fluctuation of the wall vorticity, averaged over time and the spanwise direction.
The maximum value in (\ref{eq:criteria}) is 2.79 at $x=429$, and the criterion ensures that the instantaneous wall stress at the selected location is much higher than the turbulent mean stress.
After excluding events that are already inside a developed turbulent spot, we obtain 48 events in total representing the instant when spots just hit the wall.
At later times, as a larger portion of the spot spreads along the wall, an increasing area of wall stress around each selected maximum reaches turbulent level (bottom panel in figure \ref{fig:config}) and finally merges into the fully turbulent boundary layer downstream.
Detailed information about the location and time of these events are available in Table 1 of the Supplemental Materials (SM).

The flow field around a sample event is visualized in figure \ref{fig:event}.
The stress maximum is located at $(x,y,z)=(203.5,0,220.0)$ and $t=916.25$, and the corresponding vorticity vector is $(\omega_x,\omega_y,\omega_z)=(0.78,0,-2.8)$.
The isosurfaces of $\omega_z$ in figure \ref{fig:event}$a$ resemble the ``inclined shear layers'' commonly observed in fully turbulent flows \citep{Jimenez1988}.
As speculated by \cite{Thomas_bull_1983}, ``this shear layer appears to be responsible for the characteristic variations of wall shear''.
The vortex lines in figure \ref{fig:event}$a$ are lifted above the stress minimum, and depressed towards the stress maximum at the wall (panel $b$); the pattern is reminiscent of ejection and sweep events in fully-developed wall turbulence \citep{Sheng2009}, which have been attributed to streamwise vortical structures in the buffer layer \citep{Kravchenko_Moin_1993,Orlandi_Jimenez_1994}.
In addition, fluctuations in the wall pressure gradient are appreciable during transition
(Figure \ref{fig:particles}ii), and hence the Lighthill source at early times may be important.
In fact, the contribution of the Lighthill source to the extreme events in turbulence has been controversial: 
Although the stress maximum is mostly associated with a strong pressure gradient at the wall \citep{Ghaemi_scarano_2013}, \cite{Thomas_bull_1983} claimed that the pressure pattern is not the direct cause of the bursting process, after analyzing the phase relation between the pressure and wall shear stress.
In transitional flows, whether the Lighthill source leads to the instantaneous high wall stress remains unexplored.
In the next section we quantify the contribution from each of these mechanisms using a stochastic Cauchy analysis.

\section{Results}
\label{sec:results}

We first perform a detailed analysis for the event shown in figure \ref{fig:event}, and then summarize the main results from 48 events.
A population of $10^4$ stochastic Lagrangian particles are released from the stress maximum; the instantaneous locations of 300 particles are shown in figure \ref{fig:particles}.
Once leaving the stress maximum, the particle cloud spreads immediately in all three directions due to viscous diffusion, and travels upstream due to advection by the flow velocity backward in time (figure \ref{fig:particles}$a$,$b$).
Note that near-wall particles are advected with a lower speed than the upstream propagation of the wall-stress maximum itself (panels $a$i-$c$i). Therefore, the vorticity near that maximum at earlier times cannot be the primary origin of the analyzed vorticity $\boldsymbol{\omega}(\boldsymbol x,t)$.
At $\delta s = s - t = -20$ (panel $c$), two clusters of particles can be identified: one stays near the wall (magenta), and the other (yellow) is brought towards the edge of the boundary layer by an upward velocity in reverse time. 
This pattern is consistent with the candidate mechanisms discussed in the introduction.
In forward time, the mean vertical velocity is negative at transition onset.  In addition, turbulent spots are initiated in the outer part of the boundary layer and impinge onto the wall in a ``top-down'' fashion.. 
As a result, the local vertical velocity near the release location at the wall can be positive in backward time and transport particles away from the wall.

The classification of near-wall and outer particles is performed using $k$-means clustering \citep{kmeans} based on the displacement from the release location, normalized by standard deviation of particle locations,
\begin{equation}
    \label{eq:kmeans}
    \alpha_i = \frac{\widetilde{A}^s_{t,i}(\boldsymbol x) - x_i}{\left(\widetilde{A}^s_{t,i}(\boldsymbol x) - \mathbb E[\widetilde{A}^s_{t,i}(\boldsymbol x)]\right)_{\mathrm{rms}}}, \quad i = 1,2,3.
\end{equation}
The $x$-$y$ locations of all the $10^4$ particles at $\delta s = -20$ and the classification results are reported in figure 1 of the SM.
The contribution of each cluster to the enhanced wall stress will be quantified later.
Note that most particles do not coincide with the strong $\omega_z$ region (gray isosurface in panel $c$i), which indicates that the advection of vorticity is not a dominant mechanism for the increased stress.
The spanwise Lighthill source at the wall,
\begin{equation}
    \label{eq:Lghthill_z}
    \sigma_z = \frac{\partial p}{\partial x} = -\nu \frac{\partial \omega_z}{\partial y}
\end{equation}
is plotted in figure \ref{fig:particles}ii.
Although the mean pressure gradient is zero in the simulation, the instantaneous pressure gradient at the wall is fluctuates appreciably between large positive and negative values. 
Most of the Lighthill source is concentrated within successive ``band'' structures with alternating signs (panels $a$ii, $b$ii), which are similar to the bipolar patterns commonly identified in fully turbulent flows \citep{Andreopoulos_Agui_1996,Eyink2020_channel}.
As the particles are released from the wall and strongly reflected during the initial transient, 
the unfavorable positive $\sigma_z$ region downstream of the stress maximum in panel ($a$ii) is sampled frequently.
The favorable negative $\sigma_z$ is less likely to contribute since a decreasing number of particles will revisit the wall in backward time (panels $b$ii and $c$ii).
In total, the Lighthill source is a minor, and in fact opposing, contributor to the high wall-stress in this particular event.

\begin{figure}
	\centering
	\includegraphics[width=\textwidth]{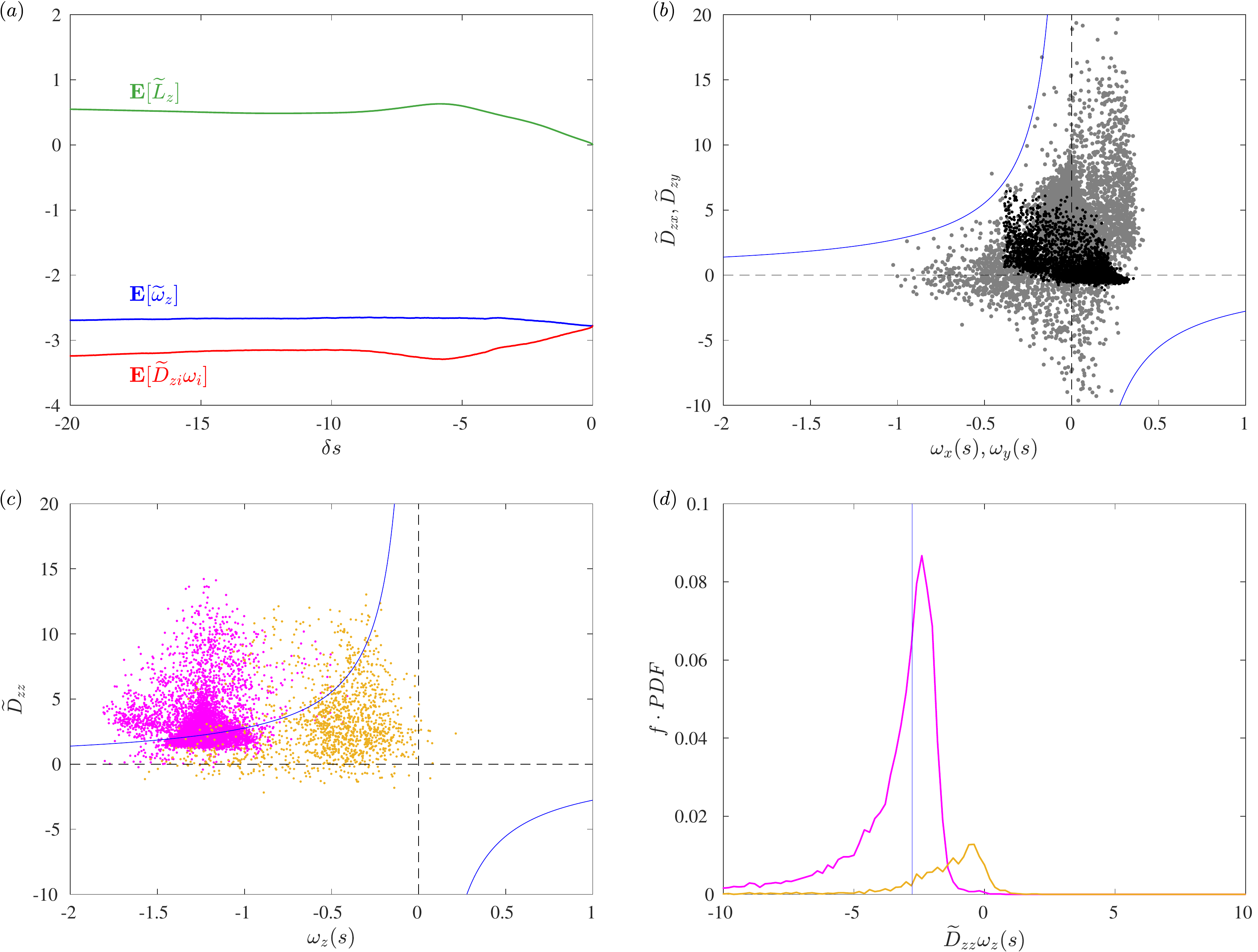}
	\caption{($a$) Temporal evolution of (blue) stochastic Cauchy invariant, (green) the contribution from Lighthill source and (red) interior deformation.
	($b$-$c$) Scatter plot of (black) ($\omega_x$,$\widetilde D_{zx}$), (gray) ($\omega_y$, $\widetilde D_{zy}$) and (magenta for near-wall cluster, yellow for outer cluster) ($\omega_z$,$\widetilde D_{zz}$) at $\delta s = -20$. Blue lines: $\widetilde D_{zi} \omega_i (s) = \omega_z(t)$, $i=1,2,3$.  
	($d$) Partial probability distribution functions (PDF's) of $\widetilde D_{zz}\omega_z(s)$ for (magenta) near-wall and (yellow) outer particles. The partial PDF's are 
	the PDF's multiplied by the fraction of particles $f$ in each cluster. Blue line marks $\widetilde D_{zz}\omega_z(s) = \omega_z(t)$.
	}
	\label{fig:stats}
\end{figure}

Since the spanwise vorticity gives rise to streamwise stress and is generally the dominant component of the wall vorticity vector, we focus on the origin of $\omega_z$. 
Expanding the expression of the stochastic Cauchy invariant (\ref{eq:SCauchy}) yields 
\begin{equation}
    \label{eq:SCauchy_z}
    \omega_z(\boldsymbol{x},t) = \mathbb{E} \left[\widetilde {\omega}_{s\,z}(\boldsymbol{x},t) \right] = \mathbb{E} \left[ \widetilde {D}_{zx}(s)\omega_{x}(s) + \widetilde {D}_{zy}(s)\omega_{y}(s) + \widetilde {D}_{zz}(s)\omega_{z}(s) + \widetilde {L}_z(s) \right],
\end{equation}
where the dependence of $\widetilde{D}$, $\omega$, $\widetilde{L}$ on target location and time $(\boldsymbol{x},t)$ has been omitted for simplicity.
The expectation of the stochastic Cauchy invariant is well conserved in backward time (blue curve in figure \ref{fig:stats}$a$), 
except for slight transient due primarily to artefacts of space-time interpolation \citep{Eyink2020_theory}, 
which confirms the theoretical analysis in \S\ref{sec:SCauchy}.
This conservation is non-trivial because the contribution to vorticity by one particle can be significantly larger than the expectation and must be cancelled by other particles such that the final Cauchy invariant is preserved.
The contribution from the Lighthill source (green in panel $a$) gradually increases
to positive values, opposite in sign to $\omega_z(t),$ due to particles sampling the wall region with $\sigma_z>0$. It then drops slightly within a short time, and remains almost a constant when $\delta s < -10$, which is the consequence of the aforementioned bipolar pattern in figure \ref{fig:particles}(ii).
The positive sign of the Lighthill source is compensated by the interior deformation with a more negative value than $\omega_z(t)$ (red in panel $a$).
In addition, the magnitude of $\mathbb{E}[\widetilde L_z]$ is less than $20\%$ of $|\omega_z(t)|$, which indicates that the wall contribution is 
not only of the wrong sign but clearly subordinate to the interior deformation.

The interior deformation at $\delta s = -20$ is further decomposed into tilting and stretching effects, presented in figures \ref{fig:stats}$b$ and \ref{fig:stats}$c$, respectively.
Each black dot in figure \ref{fig:stats}$b$ denotes the $(\widetilde D_{zx}(s),\omega_x(s))$ of one Lagrangian particle, where $\widetilde D_{zx}$ quantifies the rotation of local streamwise vorticity $\omega_x(s)$ towards the spanwise direction. 
Similarly, gray dot corresponds to tilting of wall-normal vorticity, $(\widetilde D_{zy}(s),\omega_y(s))$.
Most of the vorticity vectors $\boldsymbol \omega(s)$, especially their wall-normal component, are strongly tilted from $\delta s = -20$ to $\delta s=0$, with $\widetilde{D}_{zx},\widetilde{D}_{zy} > 1$.
Nevertheless, the importance of tilting effect is determined by the product $\widetilde{D}_{zi}\omega_{i}(s)$.
The two blue curves in figure \ref{fig:stats}$b$ mark $\widetilde D_{zi}\omega_i(s) = \omega_z(t)$, and any dots falling between these curves contribute less than the expectation of the stochastic Cauchy invariant, or equivalently, the target vorticity $\omega_z(t)$. 
Since the majority of the points in figure \ref{fig:stats}$b$ are located far from the blue curves, the tilting mechanism is insignificant to the generation of target vorticity.
Precisely, the contribution from tilting of streamwise or wall-normal vorticity is $\mathbb{E}[\widetilde{D}_{zx}\omega_x(s)]/\omega_z(t) = 1.6\%$ and $\mathbb{E}[\widetilde{D}_{zy}\omega_y(s)]/\omega_z(t) = -2.7\%$.
Therefore, spanwise stretching of the spanwise vorticity must be the dominant process inducing the enhanced wall stress, which is supported by the scatterplot of $(\omega_z(s),\widetilde{D}_{zz})$ (panel $c$) with numerous particles distributed on both sides of the blue curve marking $\widetilde D_{zz}\omega_z(s) = \omega_z(t)$.

The two clusters of points in figure \ref{fig:stats}$c$ are reminiscent of the near-wall and outer clouds in figure \ref{fig:particles}$c$.
Indeed, the left cluster consists primarily of near-wall particles (magenta), and the right cluster approximately coincides with the outer particles (yellow).
Quantitatively, the spanwise stretching term in the stochastic Cauchy invariant (\ref{eq:SCauchy_z}) can be expanded into two parts,
\begin{equation}
    \label{eq:stretch}
    \mathbb{E}\left[\widetilde{D}_{zz}\omega_z(s)\right] = f_{\mathrm{nw}} \mathbb E_{\mathrm{nw}} \left[\widetilde {D}_{zz}\omega_z(s)\right] + f_{\mathrm{out}}\mathbb E_{\mathrm{out}} \left[\widetilde{D}_{zz}\omega_z(s)\right],
\end{equation}
where $f_{\mathrm{nw}} = N_{\mathrm{nw}} / N_p $ and $f_{\mathrm{out}}= 1 - f_{\mathrm{nw}} = 1 - N_{\mathrm{nw}} / N_p $ are the fractions of near-wall and outer particles, and $\mathbb E_{(\bullet)}$ is the conditional expectation over either cluster.
Based on (\ref{eq:stretch}), the dominance of the stretching of near-wall vorticity involves two effects:
(i) $\widetilde{D}_{zz}\omega_z(s)$ of most near-wall particles concentrate around the target value $\omega_z(t)$, as shown by the conditional probability density function (PDF) in panel $d$ (magenta curve), whereas the outer cluster peaks near zero (yellow curve);
(ii) most particles belong to the near-wall cluster ($f_{\mathrm{nw}} = 86\%$), as shown by the significantly larger area under the magenta curve in panel $d$.
As a result, the enhanced skin friction is predominantly accounted for by the stretching of near-wall vorticity, $f_{\mathrm{nw}} \mathbb E_{\mathrm{nw}} \left[\widetilde{D}_{zz}\omega_z(s)\right] / \omega_z(t) = 110\%$. 
While instantaneous visualization of the flow field may show an instability growing into a spot that impinges onto the wall (c.f. Figure \ref{fig:config}), the key effect that leads to skin-friction increase is not due to the impinging turbulence transporting high-vorticity towards the wall, image vorticity or the Lighthill sources. 
Instead, hidden in these events is significant stretching of instantaneous near-wall vorticity as it advects and diffuses, and realizes the high-stress point observed on the wall.

\begin{figure}
	\centering
	\includegraphics[width=\textwidth]{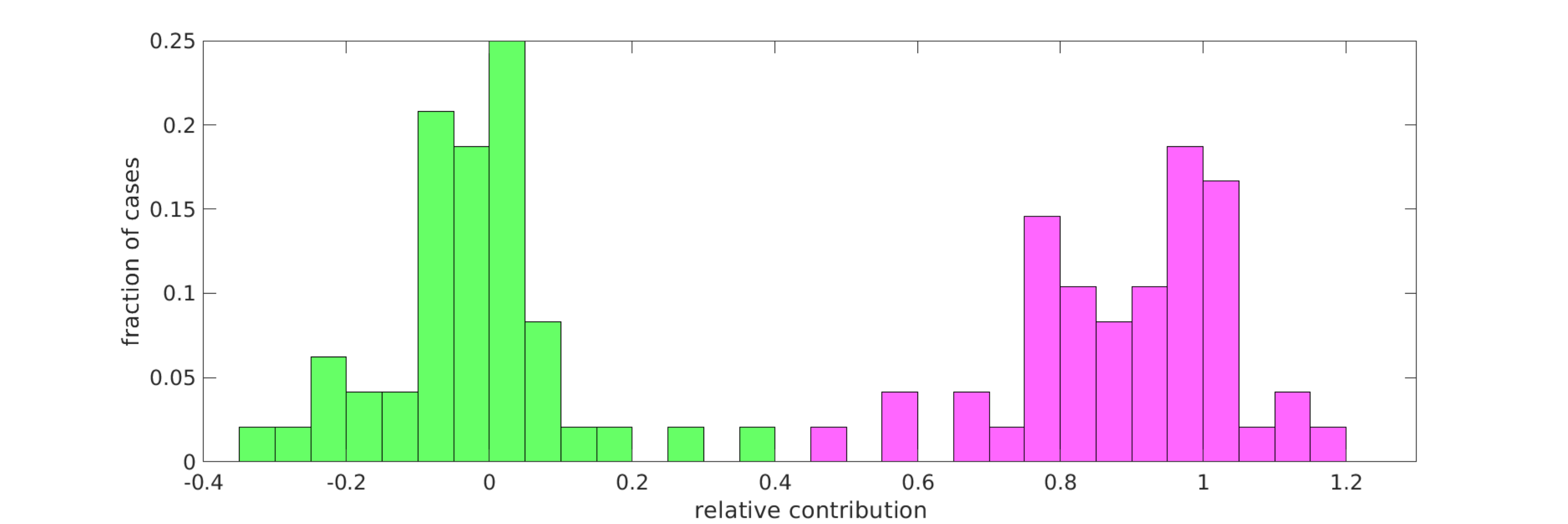}
	\caption{Histogram of the relative contribution from (green) the Lighthill source, $\mathbb E[\widetilde L_z(s)] / \omega_z(t)$, and (magenta) stretching of near-wall vorticity, $f_{\mathrm{nw}}\mathbb E_{\mathrm{nw}}[\widetilde D_{zz}\omega_z(s)] / \omega_z(t)$ for 48 events representing the enhanced skin friction.}
	\label{fig:hist}
\end{figure}

In order to examine the robustness of above conclusions against the location of time of the analyzed high wall stress, the same quantitative analysis is performed for all the 48 events obtained using the criteria in \S\ref{sec:setup}.
Specifically,
(i) the stochastic Lagrangian trajectories are integrated until $\delta s = -20$;
(ii) $k$-means clustering is implemented using the normalized displacement from the starting point (see Movie 1 in the SM for more details about the clustering results);
(iii) the expectation of different terms in the stochastic Cauchy invariant (\ref{eq:SCauchy_z}, \ref{eq:stretch}) are evaluated, 
and the results are provided in Table 1 of the SM.
The relative contribution of the Lighthill source and the stretching of near-wall vorticity are summarized in figure \ref{fig:hist}.
The Lighthill source could produce either a positive or negative contribution to the wall vorticity (green histogram), although the absolute quantity less than $40\%$ for all the cases.
By comparison, about $45\%$-$120\%$ of the enhanced skin friction originates from stretching of the near-wall vorticity (magenta histogram). 
The case with the lowest contribution from stretching has the highest contribution of the Lighthill source. 
These results confirmed the discussion from the particular event that we analyzed in detail.
Spanwise stretching of the near-wall vorticity is the dominant source to the enhanced skin friction at the onset of transition. 

Remarkably, these conclusions were essentially anticipated by \cite{Lighthill}, whose remarks deserve to be quoted here in full:
\begin{quotation}
\noindent 
``The main effect of a solid surface on turbulent vorticity close to it is to {\it correlate inflow 
towards the surface with lateral stretching}. Note that only the stretching of vortex lines can 
explain how during transition the mean wall vorticity increases as illustrated in Fig.II.21; and 
only a tendency, for vortex lines to stretch as they approach the surface and relax as they 
move away from it, can explain how the gradient of mean vorticity...is maintained in spite 
of viscous diffusion down it...\\
It is relevant to both these points that Fig.II.21 relates to uniform 
external flow, which implies zero mean rate of production of vorticity at the surface; but, even in 
an accelerating flow, the rate of production $UU'$ is too small to explain either.''
\end{quotation} 
Our exact and quantitative analysis corroborates these arguments. In particular, high magnitudes of wall vorticity are produced mainly by spanwise stretching of near-wall spanwise vorticity. We find also that the wall vorticity source makes a relatively smaller contribution. This is in part because the mean pressure-gradient of the flow is zero, so that the vorticity source is positive and negative with equal likelihood, and in part because the associated flux from the wall is too low to compete with lateral stretching. 
Even if the mean pressure gradient were not zero, as in a boundary layer with a downstream drop of total pressure, $p+\frac{1}{2}U^2,$ Lighthill in the passage quoted above argued that the average vorticity source will be too small to account for the greatly 
magnified vorticity at the wall.
This argument is not conclusive, however, because the fluctuating pressure-gradients 
may be much larger than the mean value. For example, in fully turbulent channel flow at $Re_{\tau}=1000$, \cite{Eyink2020_theory} found that the fluctuating pressure gradients 
scale in wall units as $\sim u_\tau^3/\nu$ and are larger than the mean gradient by a factor of order $Re_{\tau}$. Similar scaling is observed in the transitional flow studied here but we find, nevertheless, that the largest magnitude fluctuations of the wall vorticity source are still too small to account for the enhanced skin friction.

The vorticity-stretching mechanism highlighted by the stochastic Lagrangian analysis is complementary to previous studies based on conventional approaches \citep{Fukagata2002,Kravchenko_Moin_1993,Orlandi_Jimenez_1994}.
Near-wall structures and their dynamics are all manifest in laminar-to-turbulence transition, including e.g.\,streak instabilities, Reynolds stresses, sweeping events, and instantaneous wall pressure gradients.  
All together they provide the flow field that stretches the near-wall vorticity as it evolves from the laminar to the transitional regions of the flow, and generates the maxima in the wall stress. 
Vortex stretching above stress maxima and compression associated with stress minima lead to the formation of secondary streamwise vorticies \citep{Robinson1991}.
This iterative interaction eventually evolves into the ``cascade process'' of fully-developed turbulence \citep{Lighthill}.

\section{Discussion and conclusions}
We have explored the origin of enhanced wall friction in a transitional boundary layer, by expressing the wall stress as the expectation of a stochastic Cauchy invariant.
We proved mathematically that the expectation with Neumann boundary condition is conserved in backward time.
A Monte-Carlo scheme was adopted for numerically evaluating the invariant, and the particle trajectories were integrated by applying the Euler-Maruyama method.
Our analysis was performed using the transitional boundary layer dataset of the Johns Hopkins Turbulent Database.
We extracted 48 events of wall-stress maxima, which represent the suddenly increased skin friction at transition onset.

The invariant consists of contributions from the deformation of interior vorticity vector and the wall vorticity flux (the Lighthill source).
The effects of vortex tilting, stretching and the Lighthill source on the generation of wall vorticity were quantified and compared. 
Tilting of the streamwise or wall-normal vorticity has a small but not entirely negligible contribution.
The Lightill source can exert a favorable or adverse 
influence on the wall stress, although the relative contribution is less than 40\% for all the examined events.
Due to the upward wall-normal motion in backward time, the Lagrangian particles are separated into near-wall and outer clusters.
Among all the 48 examined events, spanwise stretching of the interior vorticity vector, especially the near-wall vorticity, is the dominant source of the enhanced skin friction, which confirmed and refined the conjecture by \cite{Lighthill}.

It is worth remarking that our generalization of the stochastic Cauchy invariant to Neumann boundary conditions
requires that the vorticity source be given by the expression in (\ref{eq:Lighthill}), 
$\boldsymbol\sigma=-\nu (\hat{\boldsymbol n} \cdot\nabla) \boldsymbol \omega|_{\mathrm w}$ which was first proposed 
for general curved walls by \cite{Panton2006}. This necessity follows from the proof in Appendix \ref{sec:proof}. 
An alternative expression proposed by \cite{Lyman1990},   $\boldsymbol\sigma=\nu \hat {\boldsymbol n} \times \left(\nabla \times \boldsymbol \omega \right)|_{\mathrm w},$  does not yield the correct result here, although it is the unique expression to describe local creation of circulation at the boundary \citep{Eyink2008}. The original work by \cite{Lighthill} considered explicitly 
only the case of a two-dimensional flat wall, where the tangential components of the two definitions agree. However, Lighthill 
assumed that the vorticity source has a non-vanishing normal component, which is only true of Panton's expression. There has been some controversy in the past over which definition of the vorticity source is ``correct'', with 
\cite{wu1996vorticity}
claiming for example that Lyman's version is inappropriate and that only Panton's expression should be used. We agree with the recent 
resolution by \cite{terrington2021generation}, which is that the two expressions measure slightly different 
things in general. Quoting directly from \cite{terrington2021generation},
``Lyman’s definition describes the transfer of circulation 
due to the tangential viscous acceleration of the fluid, while Lighthill’s definition considers only the terms that can lead to a local change in vorticity.'' This statement is consistent with our finding that the Lighthill-Panton vorticity source is the uniquely correct choice to be used as Neumann boundary condition for the 
stochastic Cauchy invariant. On the other hand, Lyman's vorticity source is continuously extended into the 
interior of the flow by the anti-symmetric vorticity flux $\Sigma_{ij}$ of \cite{huggins1994vortex}, and is 
thus related generally to pressure gradients and to energy dissipation by the Josephson-Anderson relation  \citep{Eyink2008,Eyink2021}. The generalizations to curvilinear walls of Lighthill's wall vorticity source 
by \cite{Panton2006} and by \cite{Lyman1990} have each their own proper domains of applicability, which overlap,  and one must be aware in any particular application which of the two definitions is appropriate. 

Our stochastic Cauchy analysis developed in this paper may assist in understanding various physical phenomena in transitional and turbulent wall-bounded flows. 
In addition to the strengthening of wall-vorticity, 
the reciprocal effect could also be studied of weakening of vorticity during ejection into the interior,
proposed by \cite{Lighthill} to explain the the strong concentration of vorticity near the wall in turbulent flow.
Although only bypass transition is considered in the present work, the enhanced skin friction during orderly transition may be attributed to vorticity stretching as well, since the last stage of transition is also accompanied by the formation and growth of turbulent spots.
Favorable or adverse pressure gradients might contribute to the wall vorticity during transition through the Lighthill source, but their influence is probably subordinate to vorticity stretching, as speculated by \cite{Lighthill}.
The interaction between transition and flow separation can also be interpreted from our conclusions.
Since vorticity is brought towards the wall and stretched during transition, achieving a zero-stress condition and flow separation becomes more difficult.
Therefore, control strategies that strengthen the near-wall stretching would efficiently suppress separation. 
In addition, by exploring the near-wall vorticity dynamics, the mechanism of existing drag-reduction approaches might be interpreted in a more comprehensive framework.

\appendix

\section{Numerical evaluation of the boundary local time}
\label{sec:localtime}

Recall the discrete equation for the particle location in backward time,
\begin{equation}
    \label{eq:disc_A_app}
    \widetilde{\boldsymbol{A}}_{t}^{s_{k}}(\boldsymbol{x}) = \widetilde{\boldsymbol{A}}_{t}^{s_{k-1}}(\boldsymbol{x})-\boldsymbol{u}\left(\widetilde{\boldsymbol{A}}_{t}^{s_{k-1}}(\boldsymbol{x}), s_{k-1}\right) \Delta s+\sqrt{2 \nu \Delta s} \widetilde{\boldsymbol {N}}_{k} - \nu \Delta \ell_{k} \hat{\boldsymbol y}.
\end{equation}
In this appendix, we elaborate how to evaluate the increment of the boundary local time density $\Delta \ell_k = \ell_t^{s_k} - \ell_t^{s_{k-1}}$ in the last term of (\ref{eq:disc_A_app}). 
For simplicity, we adopt a short-hand notation for $\widetilde \bA^s_t(\boldsymbol x)$,
\begin{equation}
    \widetilde \bA(s) \coloneqq \widetilde \bA^s_t(\boldsymbol x).
\end{equation}
and the equation (\ref{eq:disc_A_app}) is re-written as,
\begin{equation}
    \label{eq:localtime_A}
    \widetilde \bA(s_k) = \widetilde \bA(s_{k-1}) + \Delta \widetilde \bA(s_k;s_{k-1}) - \nu \Delta \ell_k \hat {\boldsymbol y},
\end{equation}
where at any time $s \in [s_k,s_{k-1}]$, the term $\Delta \widetilde{\boldsymbol{A}}(s;s_{k-1})$ is given by the Euler-Maruyama scheme,
\begin{equation}
    \Delta \widetilde \bA(s;s_{k-1}) = \boldsymbol u \left(\widetilde \bA(s_{k-1}), s_{k-1}\right)(s - s_{k-1}) + \sqrt{2\nu} \left(\widetilde{\boldsymbol W}(s) - \widetilde{\boldsymbol W}(s_{k-1})\right).
\end{equation}
The last term in (\ref{eq:localtime_A}) is given by the Skorohod equation \citep[chapter 2]{Zambotti2017},
\begin{equation}
    \label{eq:localtime_dl}
    \nu \Delta \ell_k=-\max \left\{0, \max _{s_{k} \leq s \leq s_{k-1}}\left\{-\widehat{\boldsymbol{y}} \cdot \left(\widetilde\bA(s_{k-1}) + \Delta \widetilde\bA(s;s_{k-1}) \right) \right\}\right\},
\end{equation}
which is non-zero only if the lowest possible location of the particle within $s \in [s_k,s_{k-1}]$ is beneath the wall.

To produce the random variable $\nu \Delta \ell_k$ with the correct statistical distribution, we can use the algorithm of \cite{Lepingle1995}.
First, the displacement without wall is evaluated, 
\begin{equation}
    \label{eq:localtime_dA}
    \Delta \widetilde \bA\left(s_{k}\right) = -\boldsymbol{u}\left(\widetilde \bA\left(s_{k-1}\right), s_{k-1}\right) \Delta s+\sqrt{2 \nu \Delta s} \widetilde {\boldsymbol N}_k,
\end{equation}
where $\widetilde {\boldsymbol N}_k$ is a standard normal random vector.
The wall-normal component of (\ref{eq:localtime_dA}), $\Delta \widetilde A_y(s_k)$, is used to evaluate the most negative displacement in $y$ (with $\triangleq$ denoting equality in distribution),
\begin{equation}
    \widetilde S_{k-1} = \max _{s_{k} \leq s \leq s_{k-1}}\left\{- \Delta \widetilde A_y(s;s_{k-1}) \right\} \triangleq \frac{1}{2}\left\{-\Delta \widetilde A_{y}\left(s_{k}\right)+\sqrt{2 \nu \widetilde V_{k-1}+\Delta \widetilde A_{y}^{2}\left(s_{k}\right)}\right\}.
\end{equation}
The exponential random variable $\widetilde V_{k-1} = -2\Delta s \ln \widetilde U_{k-1}$ is obtained from a uniform random variable $ \widetilde U_{k-1}$.
Finally, the reflected displacement from the wall is computed, 
\begin{equation}
    \nu \Delta \ell_k = -\max \left\{0, \widetilde S_{k-1}-\widetilde A_{y}\left(s_{k-1}\right)\right\},
\end{equation}
which is substituted into (\ref{eq:localtime_A}) to evaluate the particle location at time $s_k$.


\section{Conservation of the stochastic Cauchy invariant with Neumann boundary conditions}
\label{sec:proof}
In this section, we prove that the following stochastic process is a backward martingale:
\begin{equation}
    \label{eq:proof_Cauchy}
    \boldsymbol{\widetilde\omega}_s(\boldsymbol{x},t)=\widetilde\bD_{t}^{s}(\boldsymbol{x}) \cdot \boldsymbol{\omega}\left(\widetilde\bA_{t}^{s}(\boldsymbol{x}), s\right)+\int_{s}^{t} \widetilde\bD_{t}^{r}(\boldsymbol{x}) \cdot \boldsymbol{\sigma}\left(\widetilde\bA_{t}^{r}(\boldsymbol{x}), r\right) \hat{\mathrm d} \ell_{t}^{r}(\boldsymbol{x}), \quad s<t
\end{equation}
Therefore, the expectation of the stochastic Cauchy invariant (\ref{eq:SCauchy}) is conserved in backward time $s<t$.

Recall the evolution equation for the stochastic Lagrangian trajectory (\ref{eq:dAds}),
\begin{equation}
    \label{eq:proof_dAds}
    \hat{\mathrm{d}} \widetilde{\boldsymbol{A}}_{t}^{s}(\boldsymbol{x})=\boldsymbol{u}\left(\widetilde{\boldsymbol{A}}_{t}^{s}(\boldsymbol{x}), s\right) d s+\sqrt{2 \nu} \hat{\mathrm{d}} \widetilde{\boldsymbol{W}}(s) - \nu \hat{\boldsymbol n}\left(\widetilde{\boldsymbol{A}}_{t}^{s}(\boldsymbol{x}), s\right) \hat{\mathrm d} \ell_{t}^{s}(\boldsymbol{x}),
\end{equation}
the Lighthill source (\ref{eq:Lighthill}),
\begin{equation}
    \label{eq:proof_Lighthill}
    \boldsymbol \sigma = - \nu \hat{\boldsymbol n} \cdot \nabla \boldsymbol \omega|_{\mathrm{w}}
\end{equation}
and the evolution equation\eqref{Deq} for $\widetilde\bD_{t}^{s}(\boldsymbol{x})$ 
\begin{equation}
    \label{eq:proof_dD}
    d \widetilde\bD_{t}^{s}(\boldsymbol{x})=- \widetilde\bD_{t}^{s}(\boldsymbol{x}) \cdot (\boldsymbol{\nabla} \boldsymbol{u})^{\top}\left(\widetilde\bA_{t}^{s}(\boldsymbol{x}), s\right) d s.
\end{equation}
Differentiating (\ref{eq:proof_Cauchy}) with respect to $s$ and applying the product rule give
\begin{equation}
    \label{eq:proof_dCauchy}
    \hat{\mathrm d} \boldsymbol{\widetilde\omega}_s(\boldsymbol{x},t) = \widetilde\bD_{t}^{s}(\boldsymbol{x}) \cdot\left[\hat{\mathrm d} \boldsymbol{\omega}\left(\widetilde \bA_{t}^{s}(\boldsymbol{x}), s\right)-((\boldsymbol \omega \cdot \boldsymbol{\nabla}) \boldsymbol{u})\left(\widetilde\bA_{t}^{s}(\boldsymbol{x}), s\right) d s-\boldsymbol{\sigma}\left(\widetilde\bA_{t}^{s}(\boldsymbol{x}), s\right) \hat{\mathrm d} \ell_{t}^{s}(\boldsymbol{x})\right]
\end{equation}
From the backward It$\bar{{\rm o}}$ formula, one further obtains an expression for the first term in (\ref{eq:proof_dCauchy}),
\begin{equation}
\begin{aligned}
    \hat{\mathrm d} \boldsymbol{\omega}\left(\widetilde\bA_{t}^{s}(\boldsymbol{x}), s\right) &=\left(\partial_{s} \boldsymbol{\omega}+(\boldsymbol{u} \cdot \boldsymbol{\nabla}) \boldsymbol{\omega}-\nu \Delta \boldsymbol{\omega}\right) d s- \nu(\boldsymbol{n} \cdot \boldsymbol{\nabla}) \boldsymbol{\omega} \hat{\mathrm d} \ell_{t}^{s}(\boldsymbol{x})+\sqrt{2 \nu}(\hat{\mathrm d} \widetilde{\boldsymbol W}(s) \cdot \boldsymbol{\nabla}) \boldsymbol{\omega} \\
    &=((\boldsymbol{\omega} \cdot \boldsymbol{\nabla}) \boldsymbol{u})\left(\widetilde\bA_{t}^{s}(\boldsymbol{x}), s\right) d s+\boldsymbol{\sigma}\left(\widetilde\bA_{t}^{s}(\boldsymbol{x}), s\right) \hat{\mathrm d} \ell_{t}^{s}(\boldsymbol{x}) + \sqrt{2 \nu}(\hat{\mathrm d} \widetilde{\boldsymbol W}(s) \cdot \boldsymbol{\nabla}) \boldsymbol{\omega} .
    \label{eq:proof_domega}
\end{aligned}
\end{equation}
The second equality in (\ref{eq:proof_domega}) is derived based on the vorticity transport equation and the expression of the Lighthill source (\ref{eq:proof_Lighthill}).
Combining equations (\ref{eq:proof_dCauchy}) and (\ref{eq:proof_domega}) yields,
\begin{equation}
    \hat{\mathrm d} \widetilde{\boldsymbol \omega}_s(\boldsymbol x,t)=\sqrt{2 \nu} \widetilde\bD_{t}^{s}(\boldsymbol{x}) \cdot(\hat{\mathrm d} \widetilde{\boldsymbol W}(s) \cdot \boldsymbol{\nabla}) \boldsymbol{\omega}\left(\widetilde\bA_{t}^{s}(\boldsymbol{x}), s\right)
\end{equation}
which shows that $\boldsymbol{\widetilde\omega}_s(\boldsymbol{x},t)$ is a backward It$\bar{{\rm o}}$ integral and is thus a backward martingale, or a ``statistically conserved'' quantity in backward time $s<t.$

\section{Connection of the Cauchy invariant to geometric fluid mechanics}
\label{sec:geometric}

In this appendix, we briefly comment on the connection of the Cauchy invariant formulation (\S\ref{sec:CauchyReview}) and geometric fluid mechanics.  
While this connection is not essential for the contribution in the main text, it is included here for the interested reader.

The Cauchy invariant is reproduced here,
\begin{equation} 
{\boldsymbol \omega}(\ba,0)
=(\nabla_{\boldmath a}\bX(\ba,t))^{-\top} \cdot {\boldsymbol \omega}(\bX(\ba,t),t),
\tag{\ref{Cauchy-inv}} 
\end{equation}
which, as a function of the Lagrangian flow map $\bX(\ba,t)$, is a time-invariant 
quantity.  Although these invariants might appear rather trivial, they have deep geometric meaning and fully express the remarkable Lagrangian properties of vorticity for ideal fluid flows. 
As discussed at length by \cite{Besse_Frisch_2017}, the evolution equation \eqref{Lag-omega-eq}
\begin{equation} 
    \frac{d}{dt} {\boldsymbol \omega}(\bX(\ba,t),t)=
    {\boldsymbol \omega}(\bX(\ba,t),t)\cdot\nabla_{\boldmath x}\bu(\bX(\ba,t),t).
    \tag{\ref{Lag-omega-eq}}
\end{equation}
implies that vorticity is ``Lie-transported'' as a differential 2-form by the vector field of fluid velocities. 
In fact, this mathematical statement is equivalent to the fact noted in the main text that vorticity vectors are transported by the flow in the same manner as infinitesimal material line vectors. The related Cauchy formula, reproduced here
\begin{equation} 
    {\boldsymbol \omega}(\bX(\ba,t),t)
    ={\boldsymbol \omega}(\ba,0)\cdot \nabla_{\boldmath a}\bX(\ba,t), 
    \tag{\ref{Cauchy-form}}
\end{equation}
in this geometric language then appears as the exact solution of the Lie-transport equation as the ``push-forward'' of the vorticity 2-form by the Lagrangian flow map $\bX(\cdot,t)$ \citep{Besse_Frisch_2017}. 
Furthermore, the appearance of the infinitely many Lagrangian conservation laws corresponding to the Cauchy invariants is explained in the Hamiltonian formulation of the Euler equations as a consequence of an infinite-dimensional symmetry group of the action associated to relabelling of fluid particles \citep{salmon1988hamiltonian}.
An equivalent understanding arises from the geometric vision of the incompressible Euler equations by Arnold, as the equations for geodesic flow on the infinite-dimensional Lie group of volume-preserving diffeomormisms or $SDiff$ 
\citep{arnold2008topological}. 
In this context, the volume-preserving diffeomorphisms are the Lagrangian flow maps $\bX(\cdot,t)$ which satisfy 
${\rm det}(\nabla_{\boldsymbol a}\bX(\ba,t))=1$. Although long neglected, the Cauchy invariants in 
recent years have experienced a renaissance, being applied in mathematical fluid mechanics 
to establish the time-analyticity of Lagrangian particle trajectories 
\citep{zheligovsky2014time,constantin2015analyticity}, 
also to prove local existence and uniqueness
of solutions to free-surface Euler equations 
\citep{kukavica2017local}, and 
in computational fluid dynamics have yielded a novel scheme for accurate numerical solution of 
the Euler equations \citep{podvigina2016cauchy}.

 
     
  \par\bigskip
  \noindent
  \textbf{Declaration of interests.} 
  The authors report no conflict of interest.


\bibliographystyle{jfm}
\bibliography{reference}
	
\end{document}